\documentclass[aps,physrev,twocolumn,superscriptaddress]{revtex4-2}

\usepackage{graphicx}
\usepackage{lineno}
\usepackage{float}
\usepackage{amsmath}
\usepackage{amssymb}
\usepackage{babel}
\begin{document}


\title{Soliton self-excitation under pulsed driving in a Kerr resonator}


\author{Matthew Macnaughtan}
\email[]{matthew.macnaughtan@auckland.ac.nz}
\author{Zongda Li}
\author{Yiqing Xu}
\affiliation{Department of Physics, University of Auckland, Auckland, New Zealand}
\affiliation{The Dodd-Walls Centre for Photonic and Quantum Technologies, Auckland, New Zealand}
\author{Xiaoming Wei}
\author{Zhongmin Yang}
\affiliation{School of Physics and Optoelectronics, South China University of Technology, Guangzhou 510640, People’s Republic of China}
\author{St\'ephane Coen}
\author{Miro Erkintalo}
\author{Stuart G. Murdoch}
\affiliation{Department of Physics, University of Auckland, Auckland, New Zealand}
\affiliation{The Dodd-Walls Centre for Photonic and Quantum Technologies, Auckland, New Zealand}


\date{\today}

\begin{abstract}
We present a novel regime of cavity soliton excitation in a Kerr resonator driven by a train of desynchronised pulses. In this regime, the soliton solution is shown to be the sole available state for the intracavity field, allowing for the automatic excitation of single solitons without the application of any external perturbations or parameter ramping. The self-excitation of cavity soliton frequency combs is validated through numerical continuation of the Lugiato-Lefever equation, direct numerical integration, and experimental observation. We show that this regime of CS self-excitation requires only the cavity detuning and pump desynchronisation parameters to be set within the correct range, thus considerably simplifying the usually complex task of deterministic cavity soliton excitation. Additionally, we show that this procedure can also be extended to allow the deterministic generation of different families of multi-soliton bound-states. We believe this research offers a promising approach to considerably simplify cavity soliton generation in both macro- and micro-scale Kerr resonators, while also offering greatly increased thermal, power, and nonlinear efficiencies intrinsic to pulsed-driven systems.
\end{abstract}


\maketitle

\section{\label{}Introduction}

Temporal cavity solitons (CSs) are ultra-short pulses of light that can circulate indefinitely and without distortion in a coherently driven Kerr cavity. First experimentally realised in a fiber ring resonator in 2010~\cite{leo_temporal_2010}, and subsequently in optical microresonators~\cite{herr_temporal_2014}, the physics of CSs have garnered considerable attention~\cite{anderson_coexistence_2017,bao_nonlinear_2014,jang_all-optical_2016,leo_dynamics_2013,jang_ultraweak_2013}. Equally importantly, numerous real-world applications leveraging the on-chip potential of CSs and their correspondence to broadband coherent frequency combs (often referred to as soliton microcombs) have been reported, further driving interest in CS development~\cite{coddington_dual-comb_2016,trocha_ultrafast_2018,pfeifle_coherent_2014,marchand_soliton_2021,liu_photonic_2020,xu_11_2021,xu_photonic_2020,fulop_high-order_2018}. 

\begin{figure*}[ht]
    \centering\includegraphics[width=\linewidth]{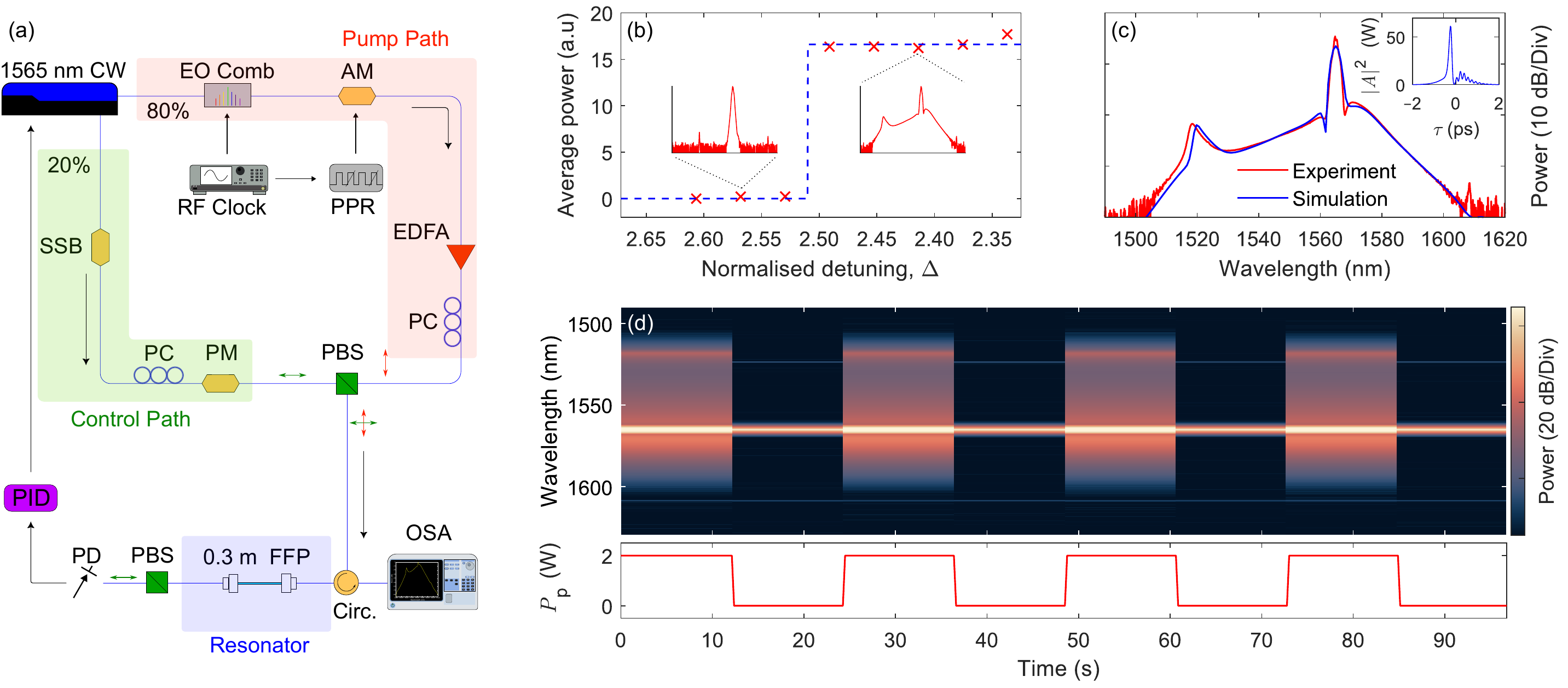}
    \caption{(a) Schematic of our experimental setup. AM: amplitude modulator, PPR: pulse picker, EDFA: erbium-doped fiber amplifier, PC: polarisation controller, PBS: polarisation beam splitter, SSB: single-sideband generator, PM: phase modulator, FFP: fiber Fabry-Perot, OSA: optical spectrum analyser, PD: photodiode, PID: proportional–integral–derivative controller. (b) Average intracavity power as the cavity detuning is scanned from high- to low-values (with the insets showing representative spectra) when the peak pump power and desynchronisation are set to 2.0~W and 2.10~fs, respectively. (c) Experimental (red) and numerical (blue) spectra of the CS state formed at $\Delta = 2.45$, with the inset showing the corresponding simulated temporal intensity profile. (d) Pseudo-coloured plot showing the experimentally acquired intracavity field spectrum as the peak pump power $P_{\text{p}}$ cycles between 2.0 W ($X=3.5$) and 0 W (shown by the line plot), with normalised cavity detuning $\Delta=2.45$ and desynchronisation $\Delta T = 2.1$~fs held constant.}
    \label{fig1}
\end{figure*}

After over a decade of scientific advancements, there has been a recent push towards more practical realisations of cavity setups and designs. This effort has primarily focused on two key practicalities of soliton comb operation: improved comb characteristics, and reliable excitation techniques. Important comb characteristics include features such as spectral bandwidth, low-noise operation, and conversion efficiency. On this front, substantial progress has been made with not only octave- and near-octave spanning microcombs~\cite{anderson_photonic_2020,okawachi_octave-spanning_2011,delhaye_octave_2011}, but also the demonstration of ultra-low noise soliton operation at quiet points~\cite{yi_single-mode_2017}, and nonlinear conversion efficiencies exceeding 40\%~\cite{xue_super-efficient_2019,xue_microresonator_2017,li_breaking_2024,li_efficiency_2022,helgason_surpassing_2023}. Significant progress has also been reported in developing new methods for CS excitation, including injection locking~\cite{Shen_turnkey_2020,wang_self-regulating_2022,lihachev_platicon_2022,voloshin_dynamics_2021,weng_turn-key_2024}, self-starting via the photorefractive effect~\cite{he_self-starting_2019}, active capture~\cite{yi_active_2016}, and chirped continuous wave (CW) driving~\cite{cole_kerr-microresonator_2018}. However, to date, there is still a shortage of platforms offering both enhanced soliton comb characteristics (i.e. broad spectral bandwidth, high conversion and power efficiency, etc.) and reliable, robust, and automatic excitation qualities. 

In this Article, we investigate a novel regime of CS excitation enabled by pulsed driving, which allows for the automatic generation of single CSs without the application of any external perturbation or parameter ramping. Furthermore, due to the use of picosecond driving pulses, our experimental scheme immediately benefits from increased power and nonlinear efficiencies, thermal stability, and a tunable repetition rate locked to an external RF clock. When operating within this self-exciting regime, the formation of a stable single CS state is guaranteed by simply setting the parameters of the drive field within a specified range. This technique thus significantly simplifies the problem of achieving reliable and deterministic CS excitation. Furthermore, we demonstrate that this technique can be extended to provide deterministic, self-exciting access to a whole family of CS bound-states~\cite{wang_universal_2017}, and that it is possible to switch back and forth between these states simply by varying either the detuning or desynchronisation of the driving field. 

\section{Experimental Setup, Numerical Model, and Demonstration of Soliton Self-excitation}

We begin by describing our experimental setup shown in Fig.~\ref{fig1}(a). Our pump field [denoted as 'pump path' in Fig.~\ref{fig1}(a)] is initially derived from a 1565~nm CW laser, which, after using standard electro-optic (EO) generation methods, is transformed into a train of 1.6~ps wide pulses~\cite{xu_frequency_2021,anderson_photonic_2020}. After passing this train of pulses through a pulse picker (PPR), the repetition rate is given as $f_{\text{rep}}$ and the pump desynchronisation (in seconds per roundtrip) can be taken to be $\Delta T = f_{\text{rep}}^{-1} - \text{FSR}^{-1}$ (where FSR represents the cavity free spectral range). Following a polarisation beam splitter (PBS) and fiber circulator, our pump field is then passed into our fiber Fabry-Perot (FFP) cavity. The cavity is formed from two partially reflective dielectric mirrors (with transmission coefficients, $\theta$, of 0.6~\%) butt-coupled to the two ends of a 0.29~m length (with round trip length $L = 0.58$~m) of dispersion-shifted fiber (DSF). The corresponding cavity FSR and finesse $\mathcal{F}$ are measured to be 350~MHz and 180, respectively. At our driving wavelength, the cavity exhibits a Kerr nonlinear coefficient of $\gamma = 2.6~\text{W}^{-1}\text{km}^{-1}$ and second-, third-, and fourth-order dispersion coefficients of $\beta_2 = -1.23~\text{ps}^2/\text{km}$, $\beta_3 = 0.13~\text{ps}^3/\text{km}$, and $\beta_4 = -5.91\times10^{-4}~\text{ps}^4/\text{km}$, respectively. To stabilise the cavity detuning, we create a separate detuning control signal [denoted as 'control path' in Fig.~\ref{fig1}(a)] which is generated from the same 1565~nm CW laser and passed through a single-sideband generator (SSB) (to shift the frequency of the control signal relative to the pump) and a phase modulator [necessary for Pound–Drever–Hall (PDH) locking]. By placing the detuning control signal on an orthogonal polarisation relative to our pump field (in which the two signals are combined into one path through a PBS), we can easily lock our driving laser to the control signal [which we detect via a slow photodiode (PD)] with a proportional–integral–derivative (PID) controller, thus stabilising the linear cavity detuning and allowing us to tune it freely.~\cite{li_experimental_2020,li_ultrashort_2024,coen_experimental_1998}.

To numerically model our system we use the well-known mean-field Lugiato-Lefever equation (LLE)~\cite{coen_modeling_2013,hendry_impact_2019}:
\begin{widetext}
\begin{equation}\label{LLE}
    t_{\text{R}}\frac{\partial A(\tau,t)}{\partial t} =  i\gamma L\Big[ (1-f_{\text{R}})|A|^{2} + f_{\text{R}}h_{\text{R}}(\tau)\circledast|A|^{2}  \Big]A 
	+ \Big[-\alpha -i\delta +  iL \sum_{k=2}^{4} \frac{\beta_{k}}{k!} \Big( i\frac{\partial}{\partial \tau}\Big)^k  -\Delta T\frac{\partial}{\partial \tau} \Big]A + \sqrt{\theta} A_{\text{in}}(\tau) ,
\end{equation}
\end{widetext}
Here, $A(\tau,t)$ and $t_{\text{R}}$ represent the intracavity field envelope and cavity roundtrip time, with $\tau$ (describing how the intracavity field evolves in a reference frame travelling at the group velocity of the driving field) and $t$ (describing how the intracavity field evolves over consecutive roundtrips) being the fast- and slow-time variables, respectively. The first term in brackets models the nonlinear interactions within the cavity, and includes both the instantaneous Kerr response and delayed Raman response $h_{\text{R}}(\tau)$ (which we model using a multi-vibrational model~\cite{hollenbeck_multiple-vibrational-mode_2002}) with Raman fraction $f_{\text{R}} = 0.18$. The second term in brackets includes the cavity loss $\alpha$ (where $\alpha = \pi/\mathcal{F}$), linear detuning, chromatic dispersion, and pump desynchronisation. The final term describes the driving field and is related to the pump power by $P(\tau) = |A_{\text{in}}(\tau)|^2$. Some useful measures that we will use to facilitate comparison throughout this Article are the nominalised pump power $X = P\gamma L \theta / \alpha^3$ and detuning $\Delta = \delta/\alpha$. 

To demonstrate soliton self-excitation experimentally, we first conduct a detuning ramp from high- to low-values while driving the cavity with a train of 2.0~W peak power pulses ($X = 3.5$) at a pump desynchronisation of $\Delta T = 2.10$~fs. Figure~\ref{fig1}(b) shows how the average intracavity power changes abruptly as we vary the detuning, showing that after the detuning passes below some critical value, the intracavity field appears to spontaneously and unexpectedly switch from a low-energy steady-state into some stable solitonic structure (an example spectrum of which is shown in the inset). Figure~\ref{fig1}(c) shows a single experimental spectral trace in red of the intracavity field seen in (b) when the normalised detuning is $\Delta = 2.45$, with the overlaid blue trace showing a numerically simulated spectrum (with the temporal intensity profile corresponding to the simulated spectrum shown in the inset) obtained using the same experimental parameters. Along with the excellent agreement between experiment and simulation that confirms the presence of an ultra-short CS, we note the presence of a strong phase-matched spectral peak around 1520~nm due to the non-negligible higher-order dispersion (HOD)~\cite{li_experimental_2020,macnaughtan_temporal_2023,jang_observation_2014}. By stabilising the intracavity field just beyond the critical detuning seen in (b), we then observe [see Fig.~\ref{fig1}(d)] the intracavity field spontaneously forming from an empty cavity into a stable CS structure as we periodically turn our driving field off and on. Within this regime, the CS appears to act as the sole attractor state of the cavity, forcing the intracavity field to evolve into a CS regardless of the initial conditions.

The soliton self-excitation observed in Figs.~\ref{fig1}(b) and (d) sits in stark contrast to more conventional CW-driven systems in three key features. Firstly, under CW-driving, following a perturbation to the intracavity field like that seen in Fig.~\ref{fig1}(d) we would expect the system to evolve towards the lower homogenous steady-state (HSS). Secondly, CS formation via a detuning scan can generally only occur when the detuning is ramped from low- to high-values, oppositive to what we show in Fig.~\ref{fig1}(b)~\cite{coen_temporal_2015,li_observations_2021}. Finally, CS generation, in our case, is a fully deterministic process, whereas in CW systems, adiabatic scanning of a control parameter (e.g. detuning) is typically stochastic and leads to a random configuration of CSs~\cite{karpov_dynamics_2019}. 

\begin{figure}[b]
    \centering\includegraphics[width=\linewidth]{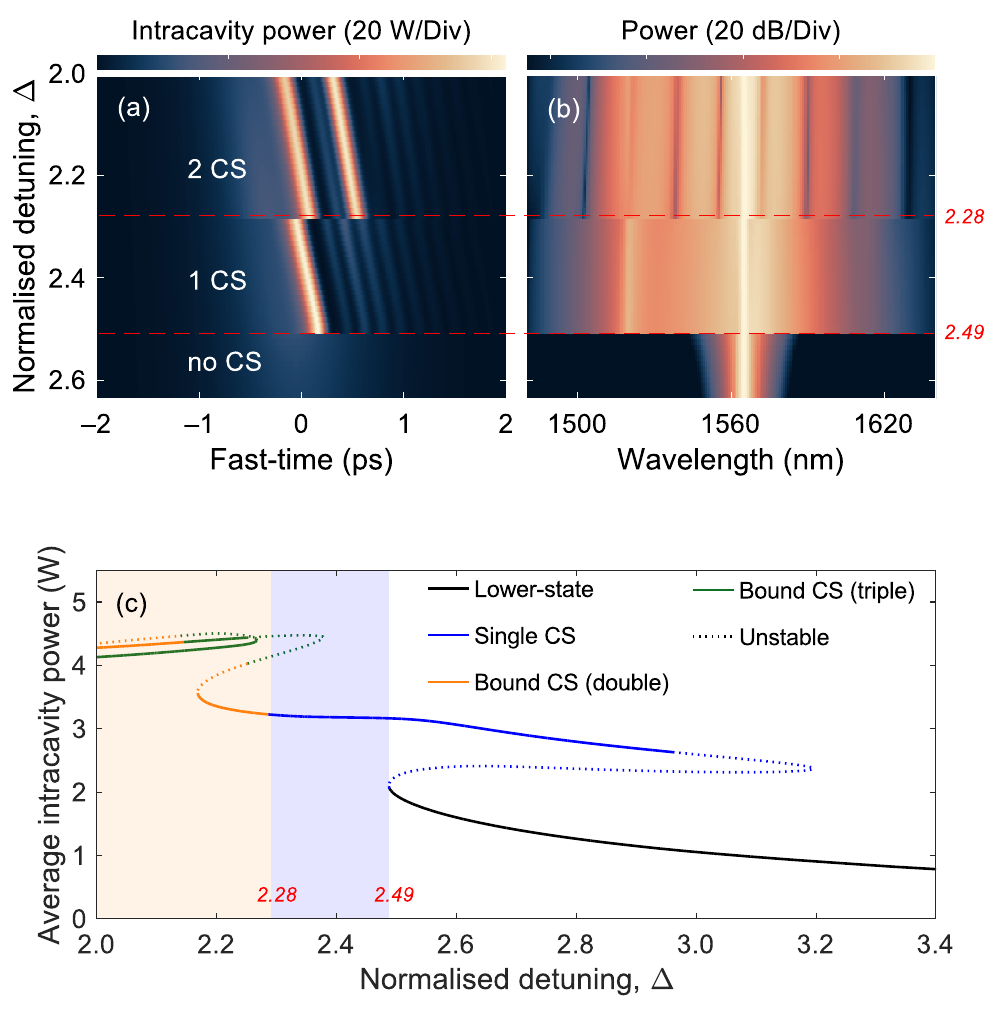}
    \caption{(a) and (b) show numerically simulated temporal and spectral intensity profiles obtained via numerical split-step integration of~\eqref{LLE} for different values of the cavity detuning [with the same cavity and pump parameters as in Fig.~\ref{fig1}]. (c) Newton-Raphson continuation using the same parameters as in (a) and (b). The shaded blue (orange) region shows where single (bound) CSs are the only states available, and thus will be self-excited. The black, blue, orange, and green solid curves represent the HSS, single CS, and double- and triple-bound CS states, respectively, with the dotted curves showing the unstable states.}
    \label{fig2}
\end{figure}

\section{Theory of soliton self-excitation}

To further explore CS self-excitation, we conduct a number of separate numerical simulations of \eqref{LLE} over a range of detunings [with the same cavity and pump pulse parameters seen in Fig.~\ref{fig1}]. Each simulation is initiated from an empty cavity, with an initial condition corresponding to broadband noise. Figures~\ref{fig2}(a) and (b) show the temporal and spectral steady-state profiles at each detuning. In agreement with the experimental results shown in Fig.~\ref{fig1}, we find a range of detunings in which the intracavity field spontaneously shapes into a stable CS. Furthermore, the numerically predicted critical detuning ($\Delta = 2.49$) that marks the beginning of the self-excitation regime seen in Figs.~\ref{fig2}(a) and (b) is in good agreement with the experimental measurements seen in Fig.~\ref{fig1}(b). We can also identify a sizeable region of detunings where CS bound-states are self-excited, as evidenced by the two bright CS structures and modulated spectral envelope in Figs.~\ref{fig2}(a) and (b), respectively. 

To gain more insights, we use a Newton-Raphson solver to find the stable and unstable steady-state solutions of~\eqref{LLE} under pulsed-driving conditions. This procedure not only illuminates the available cavity states and their transitions, but also informs us of their mutual coexistence or lack thereof. Using the same cavity and driving parameters as in Figs.~\ref{fig2}(a) and (b), Fig.~\ref{fig2}(c) presents results from a numerical continuation in detuning. Here, the solid (dotted) curves indicate stable (unstable) states, the blue curve shows the single soliton state, the orange and green curves show bound double-CS and triple-CS states, and the black curve shows the low-intensity steady-state (analogous to the lower-HSS). The shaded blue and orange regions indicate where single and bound CS self-excitation is observed in (a) and (b).

\begin{figure}[t]
    \centering\includegraphics[width=\linewidth]{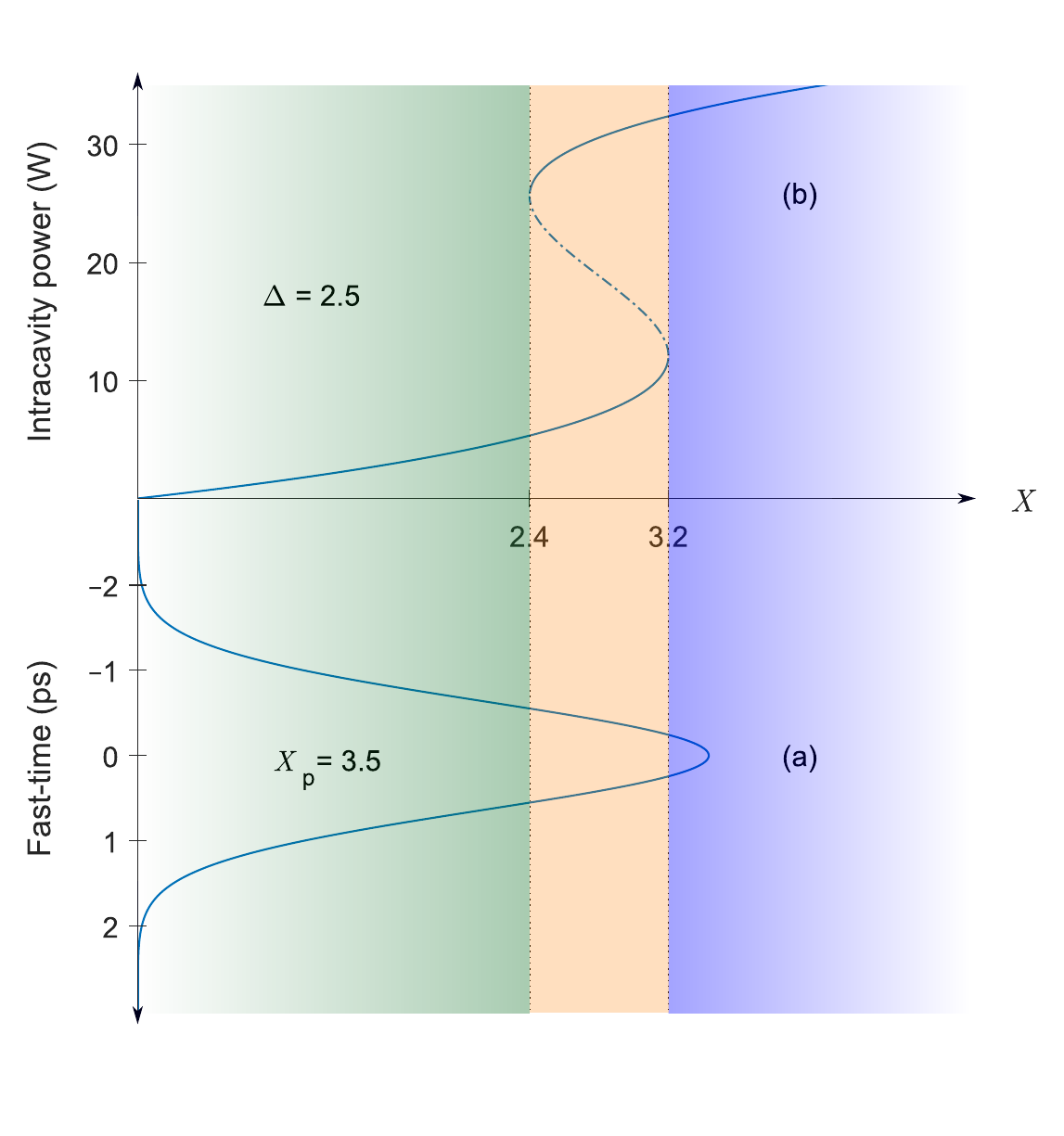}
    \caption{(a) Temporal profile of a 1.6~ps wide drive pulse with a peak normalised power of $X_{\text{p}} = 3.5$. The blue, orange, and green sections highlight the pump powers in which the local stability (when the normalised detuning is set to $\Delta = 2.5$) maps to the upper-HSS, bistable, and lower-HSS regions of the equivalent CW cavity response, respectively. (b) S-shaped CW response with the blue, orange, and green regions corresponding to the same stability regions indicated in (a).}
    \label{fig3}
\end{figure}

We recall that CSs within conventional CW-driven systems exist on a hysteresis curve (terminating at the up-switching point~\cite{coen_universal_2013}) and necessarily coexist with the lower-HSS. This means that if the initial intracavity field is in some low-power state, such as an empty cavity, then without a significant external perturbation, the cavity will evolve towards the lower-HSS. In stark contrast, once the normalised detuning is set below 2.49 in our pulsed-driven system, the lower-HSS ceases to exist, leaving only the CS state (blue shaded area). As a result, when the pump detuning is set within this regime, the intracavity field will necessarily and automatically form into a stable CS [as seen experimentally in Figs.~\ref{fig1}(b) and (d)]. We note that although HOD, Raman scattering, and pump desynchronisation were included in this continuation to match best the experimental results seen in Fig.~\ref{fig1}, the presence of these higher-order effects are not critical to the existence of CS self-excitation. Indeed, simulations in a pure Kerr cavity --- with only second-order dispersion and no pulse desynchronisation --- show a very similar behaviour.

\begin{figure}[b]
    \centering\includegraphics[width=\linewidth]{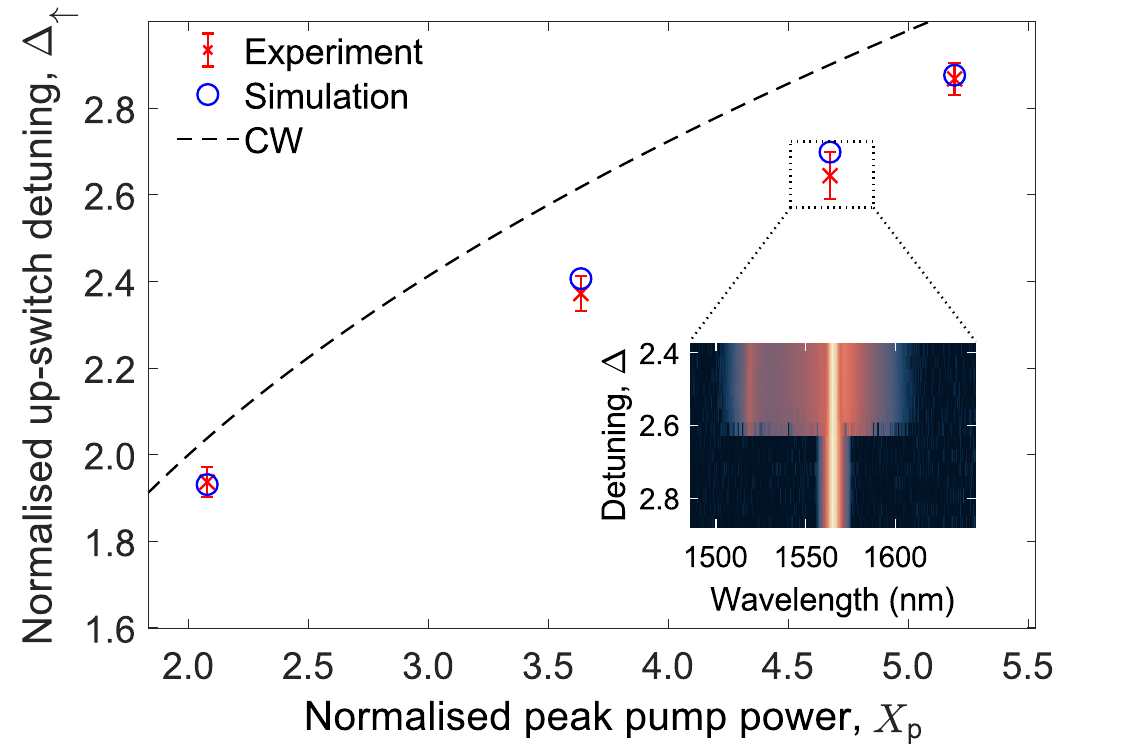}
    \caption{Maximum detunings for CS self-excitation observed experimentally (red crosses) and numerically (blue circles) at varying peak pump powers. The inset illustrates an example measurement and shows a pseudo-coloured plot of the intracavity spectrum. The dashed black line shows the theoretically predicted CW up-switching point.}
    \label{fig4}
\end{figure}

\begin{figure*}[ht]
    \centering\includegraphics[width=\linewidth]{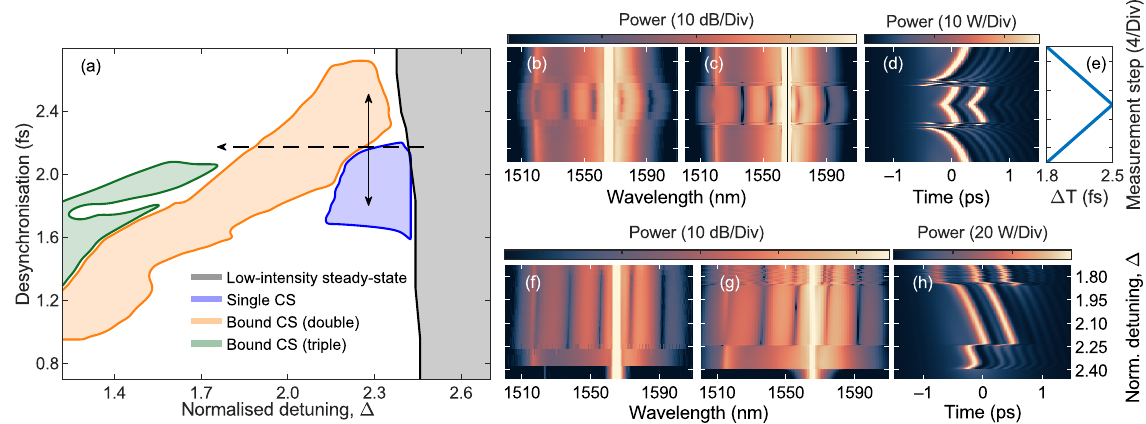}
    \caption{(a) Existence regions of both single- (blue), double- (orange), and triple-CS (green) states that exhibit self-excitation when driven with a train of 1.6 ps wide pulses with $P_{\text{p}} = 1.85$~W ($X = 3.20$). (b) Pseudo-coloured plot showing the experimental evolution of the spectrum of the intracavity field as the desynchronisation is scanned back and forth [denoted by the double-sided solid arrow in (a) and shown in (e)]. (c) and (d) show numerically simulated spectral and temporal profiles, respectively, while under conditions similar to the experiments in (b). (f) shows an experimentally acquired reverse detuning scan corresponding to the dashed arrow in (a), while (g) and (h) show the equivalent numerically simulated spectral and temporal profiles, respectively.}
    \label{fig5}
\end{figure*}

The lack of a low-power steady-state below a normalised detuning of 2.49 can be qualitatively explained by examining the bistable response of a pulse-driven cavity. In the absence of dispersion and desynchronisation, each position along the intracavity field can be approximated as independent elements driven at a pump power at the corresponding temporal position, meaning that the intracavity response varies along the driving pulse (see Fig.~\ref{fig3}). For parameters within the self-excitation regime, a small region around the peak of the driving pulse maps to the upper-HSS [blue region in Figs.~\ref{fig3}(a) and (b)], while regions away from the centre, with lower drive powers, map to either bistability [orange region in Figs.~\ref{fig3}(a) and (b)] or the lower-HSS [green region in Figs.~\ref{fig3}(a) and (b)]. Consequently, across the drive pulse’s temporal profile, there must necessarily be a transition of the intracavity field from the lower- to upper-HSS~\cite{garbin_experimental_2017,macnaughtan_temporal_2023,lottes_excitation_2021}. This enforced switching acts as the seed for CSs, allowing the soliton to self-start in a cavity operated under these conditions. We note that this mechanism is not only qualitatively similar to CS generation under chirped CW driving~\cite{cole_kerr-microresonator_2018}, but also provides further insight into previously studied spontaneous temporal symmetry breaking in pulse-driven cavities~\cite{Xu_symmetry_2014}, of which our system represents a generalised case.

The above analysis allows us to estimate an upper bound for the critical detuning at which CS self-excitation can occur as the up-switching point $\Delta_{\uparrow}(P_{\text{p}})$ of a homogeneously CW-driven cavity with a pump power that coincides with the peak pump power $P_{\text{p}}$ of our driving pulses~\cite{haelterman_dissipative_1992,coen_universal_2013}. Since this quasi-CW analysis does not take into account the additional effects of dispersion and desynchronisation, we find that, in practice, this prediction slightly overestimates the actual maximum detuning for self-CS excitation. In Fig.~\ref{fig4}, we plot the maximum detuning at which CS self-excitation occurs. The black dashed line shows the theoretical prediction from the quasi-CW analysis described above, while red crosses (blue circles) show the experimentally (numerically) observed critical detuning. Overall, we see good agreement between experiment and simulation, with the simplified CW analysis providing a useful upper bound for this phenomenon. 

\section{Evolution of self-excitation with Driving parameters}

Our numerical and experimental observations indicate that CS self-excitation dynamics are fundamentally governed by the pulse-driving configuration. This naturally raises the question of how these dynamics depend on key, easily tuneable driving parameters such as the desynchronisation, detuning, and pump power. Furthermore, how can we optimise these parameters to facilitate more practical and efficient CS self-excitation. To explore the full parameter space that permits self-excitation, we repeat the numerical simulations presented in Fig.~\ref{fig2} but over a broad range of desynchronisations. Figure~\ref{fig5}(a) shows a numerically simulated self-excited soliton existence map as the normalised detuning and desynchronisation of the cavity are varied when driven with a 1.6~ps wide pump pulse at a peak power of  $P_{\text{p}} = 1.85$~W ($X = 3.20$). The blue region represents parameter values where single self-excited CSs are supported, while the orange and green regions indicate stable double- and triple-CS bound states, respectively. The white area corresponds to unstable operation, and the black region represents the low-intensity steady-state. Notably, there is a broad range of parameters that allow both single- and bound-CS states to self-excite, demonstrating a high tolerance for variations in detuning and desynchronisation. This broad operational range simplifies experimental access to CS self-excitation under pulsed-driving, potentially reducing equipment costs by making it feasible to use more affordable electronic components (e.g. RF synthesizers).

Examining Fig.~\ref{fig5}(a), we see that the single- and double-CS regions are almost contiguous to each other, with only a small region of instability separating the two. Thus, deterministic transitions between the two states should be possible by following a desynchronisation path such as the one indicated by the double-sided solid arrow seen in (a). Figure~\ref{fig5}(b) shows experimentally acquired spectra of the cavity output as we adjust the desynchronisation of our cavity to follow the path shown in Fig.~\ref{fig5}(e). As predicted, this results in clear transitions between the single and double CS states, each readily identified by their characteristic spectral interference patterns. Repeating this path allows us to cross back and forth between these two states. This is explored further in Figs.~\ref{fig5}(c) and (d), which show the spectral and temporal cavity output obtained from a numerical simulation that follows the same desynchronisation path, showing overall excellent agreement between experiment and simulation.

We can also perform a reverse ramp in detuning indicated by the dashed arrow in Fig.~\ref{fig5}(a), with Fig.~\ref{fig5}(f) presenting the experimentally acquired spectra for this parameter scan. As predicted, clear transitions between the low-intensity steady-state, stable single- and double-CS states, and unstable states are observed. These observations are further corroborated in Figs.~\ref{fig5}(g) and (h), which show numerically simulated spectral and temporal intensity profiles along the same parameter path used in Fig.~\ref{fig5}(f). The agreement between the experimental and numerical simulations is very good, demonstrating the high degree of control achievable within this system.

\begin{figure}[t]
    \centering\includegraphics[width=\linewidth]{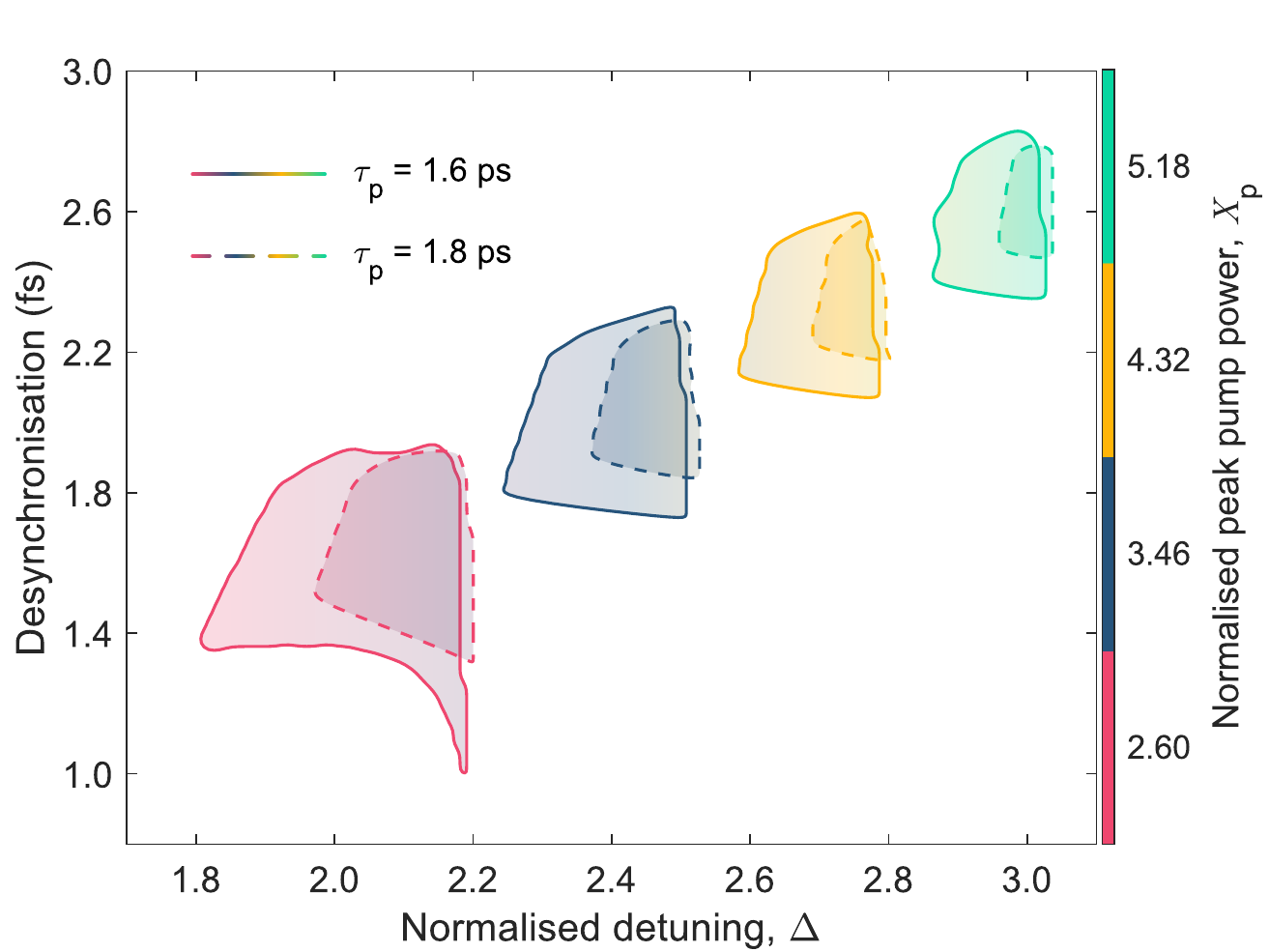}
    \caption{Numerically simulated existence boundaries of single-CS states that exhibit self-excitation are shown for a range of normalised peak pump powers $X_{\text{p}}$. The boundaries are plotted for two driving pulse widths: $\tau_{\text{p}} = 1.6$~ps (solid curves) and  $\tau_{\text{p}} = 1.8$~ps (dashed curves).}
    \label{fig6}
\end{figure}

Finally, Fig.~\ref{fig6} presents the parameter boundaries of single CS self-excitation for a range of normalised peak pump powers $X_{\text{p}}$ and pump pulse durations $\tau_{\text{p}}$ through numerical simulation of the LLE. Overall, we see that as the peak pump power is increased, the region of single CS self-excitation is pushed to higher detunings and desynchronisations. In the case of detuning, this tendency is expected, as the analysis in the previous section showed that an upper bound for this regime is given by the up-switching point, which itself increases with pump power. For desynchronisation, the increasing trend can be attributed to the growing group-velocity mismatch between the CS and pump field as the pump power rises. Interestingly, as both the pump power and pump pulse duration increase, the parameter space for single CS self-excitation noticeably shrinks. Regarding pump power, this behaviour mirrors that observed in conventional CW-driven systems, where higher pump powers tend to induce greater instabilities~\cite{leo_dynamics_2013,karpov_dynamics_2019}. As for the pump pulse duration, the effect can be understood by recognizing that as the pump pulse duration increases, the cavity dynamics gradually approach those of a conventional CW-driven system, ultimately leading to the disappearance of CS self-excitation. Additionally, we note that as the pump pulse duration increases, the onset of CS self-excitation at each pump power appears to be pushed to higher detunings, i.e. closer to the up-switching point. This occurs due to the localised switching responsible for CS self-excitation; specifically, as the pump pulse duration is increased, a larger region of the driving pulse will exclusively exist within the upper-HSS (for a detuning before the up-switching point). In essence, broader driving pulses provide a larger region for the intracavity field to switch up into a CS state at detunings closer to the up-switching point. 

Considering Figs.~\ref{fig5} and \ref{fig6} holistically, we observe that single CS self-excitation is maximized at lower peak pump powers and shorter pump pulse durations. Moreover, the parameter landscape is populated with higher-order bound-state self-exciting structures that can be easily and deterministically tuned into.

\section{Conclusion}

To conclude, we have both demonstrated and explored CS self-excitation via a pulsed-driving source within a passive Kerr resonator. Furthermore, this approach enables the deterministic selection of various CS bound states due to the interplay between the localized driving field and cavity dynamics. Operating within this self-exciting regime provides a robust method of CS generation that does not require any external perturbation or parameter ramping while leveraging the enhanced thermal, power, and nonlinear efficiencies inherent to pulsed-driven systems.

Our investigation revealed that, within the self-exciting regime, CS states act as unique attractors, directing the intracavity field towards soliton solutions regardless of the initial conditions (such as an empty cavity). This self-excitation process is driven by self-switching, arising from variations in local CW stability across the driving pulse, a mechanism intrinsically linked to pulsed-driving. Furthermore, our experimental and numerical exploration of the parameter space shows that CS self-excitation is not confined to a narrow region but is, in fact, extensive and easily accessible with standard system parameters, such as the cavity detuning and pump desynchronisation. This robustness opens the door to integrating more accessible electrical components, potentially reducing system costs precipitously.

We have also identified that the regime for single CS self-excitation is most pronounced at lower pump powers (having observed self-excitation over a normalised drive power range of $X$~=~2~--~6) and shorter pump pulse durations. This suggests a strong compatibility with microresonator platforms similar to those reported in Refs. \cite{obrzud_temporal_2017,cole_kerr-microresonator_2018,qureshi_soliton_2022}. Additionally, with the rapid advancements in integrated photonic devices, particularly in electro-optic crystal-based platforms~\cite{zhang_power_efficient_2023,kuznetsov_ultra_broadband_2025,zhang_ultrabroadband_2025}, there is significant potential for a fully integrated microresonator system under electro-optic pumping that utilises CS self-excitation.

Overall, this work highlights the previously under-explored nonlinear dynamics of CS self-excitation under pulsed driving and provides a robust, efficient, and accessible method for CS generation in both macro- and microresonator systems.

\bibliography{ref2}

\end{document}